\documentclass[3p,times,procedia]{elsarticle}
\usepackage{nupha_ecrc}

\volume{00}

\firstpage{1}

\journalname{Nuclear Physics A}

\runauth{}

\jid{nupha}

\jnltitlelogo{Nuclear Physics A}

\usepackage{float}

\usepackage{amssymb}
\usepackage{amsmath}

\usepackage[figuresright]{rotating}

\begin{document}

\begin{frontmatter}



\dochead{XXVIIth International Conference on Ultrarelativistic Nucleus-Nucleus Collisions\\ (Quark Matter 2018)}

\title{Chiral phase transition of (2+1)-flavor QCD}



\author[CCNU]{H.-T.\ Ding}
\author[IISc]{P.\ Hegde}
\author[BI,BNL]{F.\ Karsch}
\author[BI]{A.\ Lahiri}
\author[CCNU]{S.-T.\ Li}
\author[BNL]{\\ S.\ Mukherjee}
\author[BNL]{P.\ Petreczky}
\address[CCNU]{Key Laboratory of Quark \& Lepton Physics (MOE) and Institute of Particle Physics, \\
Central China Normal University, Wuhan 430079, China}
\address[IISc]{Center for High Energy Physics, Indian Institute of Science, Bangalore 560012, India}
\address[BI]{Fakult\"{a}t f\"{u}r Physik, Universit\"{a}t Bielefeld, D-33615 Bielefeld, Germany}
\address[BNL]{Physics Department, Brookhaven National Laboratory, Upton, New York 11973, USA}

\begin{abstract}
We present here results on the determination of the critical temperature in the chiral limit
for (2+1)-flavor QCD. We propose two novel estimators of the chiral critical temperature where
quark mass dependence is strongly suppressed compared to the conventional estimator using
pseudo-critical temperatures. We have used the HISQ/tree action for the numerical simulation
with lattices with three different temporal extent $N_{\tau}=$6, 8, 12 and varied
the aspect ratio over the range $4 \leq N_{\sigma}/N_{\tau} \leq 8$. To approach the chiral
limit, the light quark mass has been decreased keeping the strange quark mass fixed at its
physical value. Our simulations correspond to the range of pion masses,
55 MeV $\leq m_{\pi} \leq$ 160 MeV.
\end{abstract}


\end{frontmatter}


\section{Introduction}
\label{sc.intro}
\vspace*{-0.3 cm}

Strongly interacting matter under extreme conditions undergoes a transition from a chirally
broken (confined) phase to a chirally restored (deconfined) phase at a certain temperature.
At vanishing baryon density, this transition is a crossover for physical pion mass and the
corresponding pseudo-critical crossover temperature has been determined \cite{HotQCDTpc12}
with very good accuracy \cite{PSQM18}. In the chiral limit, {\it i.e.}\ in the limit of vanishing
light quark mass, two possible scenarios have been conjectured, the chiral phase transition
can either be (1) of $2^{\rm nd}$ order belonging to $O(4)$ universality class \cite{KMSetal82,PW84}
or to $U(2)\times U(2)$ universality class \cite{PV13} or (2) it can be of $1^{\rm st}$
order \cite{PW84}. In the later case, a critical light quark mass will exist, at which the transition
will be of $2^{\rm nd}$ order belonging to $Z(2)$ universality class. In this work our main goal
is to determine the chiral transition temperature, $T_c^0$, and we will shed some light on the
nature of the chiral transition.

\vspace*{-0.5 cm}
\section{Observables and definitions}
\label{sc.obsanddef}
\vspace*{-0.3 cm}

The quark condensate per flavor $f$ is defined as 
$\langle\bar{\psi}\psi\rangle_f=TV^{-1}\partial \ln Z\left(T,V,\{ m_f \}\right)/\partial m_f$.
The light quark condensate $\langle\bar{\psi}\psi\rangle_l$ is the order parameter of the chiral
phase transition. It requires additive as well as multiplicative renormalizations. We therefore 
have worked with a renormalization group invariant quantity, 
$M=2 \left(m_s \langle\bar{\psi}\psi\rangle_l - m_l \langle\bar{\psi}\psi\rangle_s \right)/f_K^4$,
where we have used $f_K=\left(156.1/\sqrt{2}\right)$ MeV to set the scale.
The corresponding susceptibility is defined as $\chi_{M}=m_s\partial M/\partial m_l$,
where $m_l$ and $m_s$ are light and strange quark masses, respectively.

Near the critical point the order parameter $M$ and its susceptibility $\chi_{M}$ are expected to behave like
$M= h^{1/\delta}f_G(z) + f_{\rm sub}(T,H)$ and
$\chi_{M}= h^{-1}_0 h^{1/\delta-1}f_{\chi}(z) + \partial f_{\rm sub}(T,H)/\partial H$, where $f_G(z)$
and $f_{\chi}(z)$ are universal scaling functions, which have been determined previously from spin model
calculations, and $f_{\rm sub}(T,H)$ takes into account corrections-to-scaling terms and regular terms.
The scaling variable $z$ is defined as
$z=th^{-1/\beta\delta}=z_0\left(\left(T-T_c^0\right)/T_c^0\right) H^{-1/{\beta\delta}}$
with $t=t_0^{-1}\left(T-T_c^0\right)/T_c^0$ and $h=(m_l/m_s)/h_0=H/h_0$. The scale $z_0$ is defined
as $z_0=h_0^{1/\beta\delta}/t_0$
with $T_c^0$ being the chiral critical temperature and $H$ being the symmetry breaking field.

\begin{figure}[H]
\begin{center}
\includegraphics[scale=0.50]{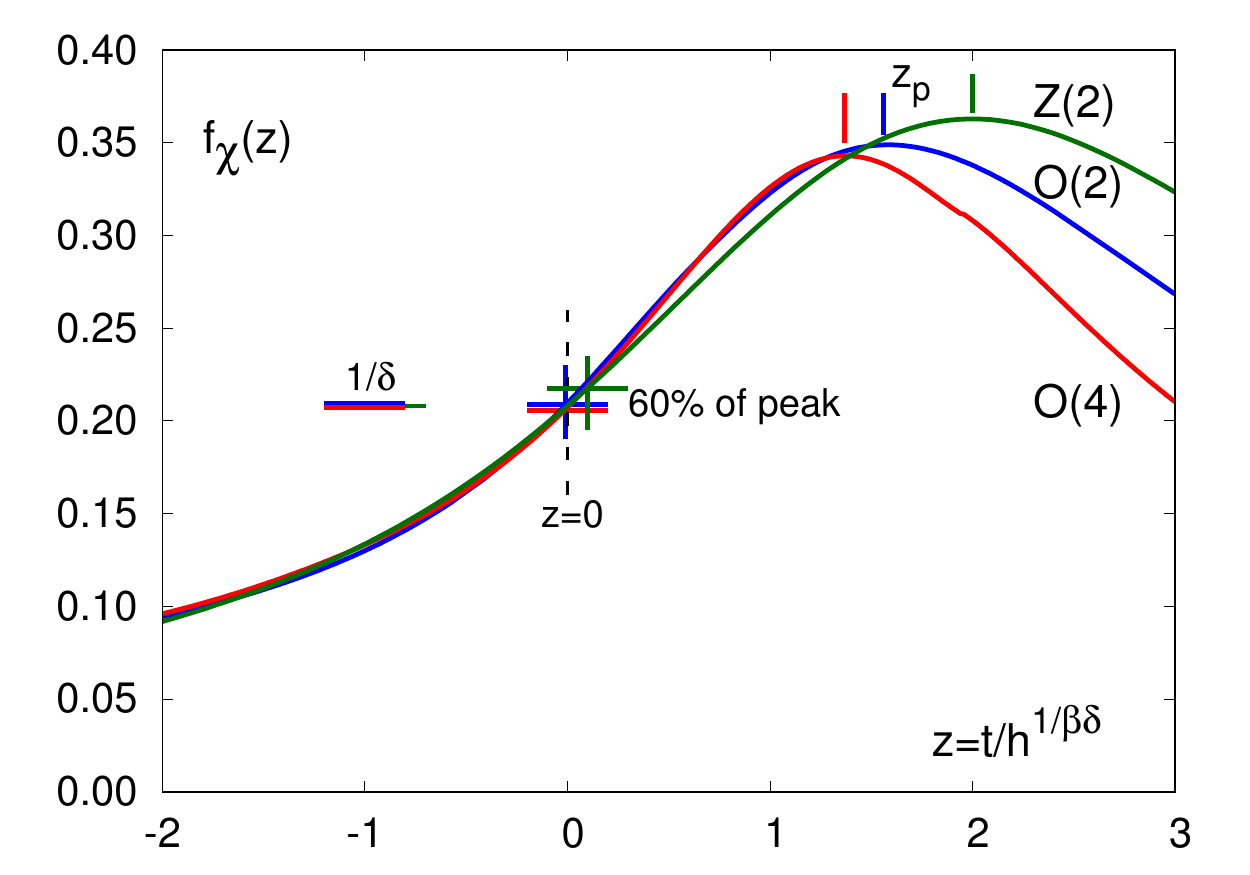} ~~~~~~~~
\includegraphics[scale=0.50]{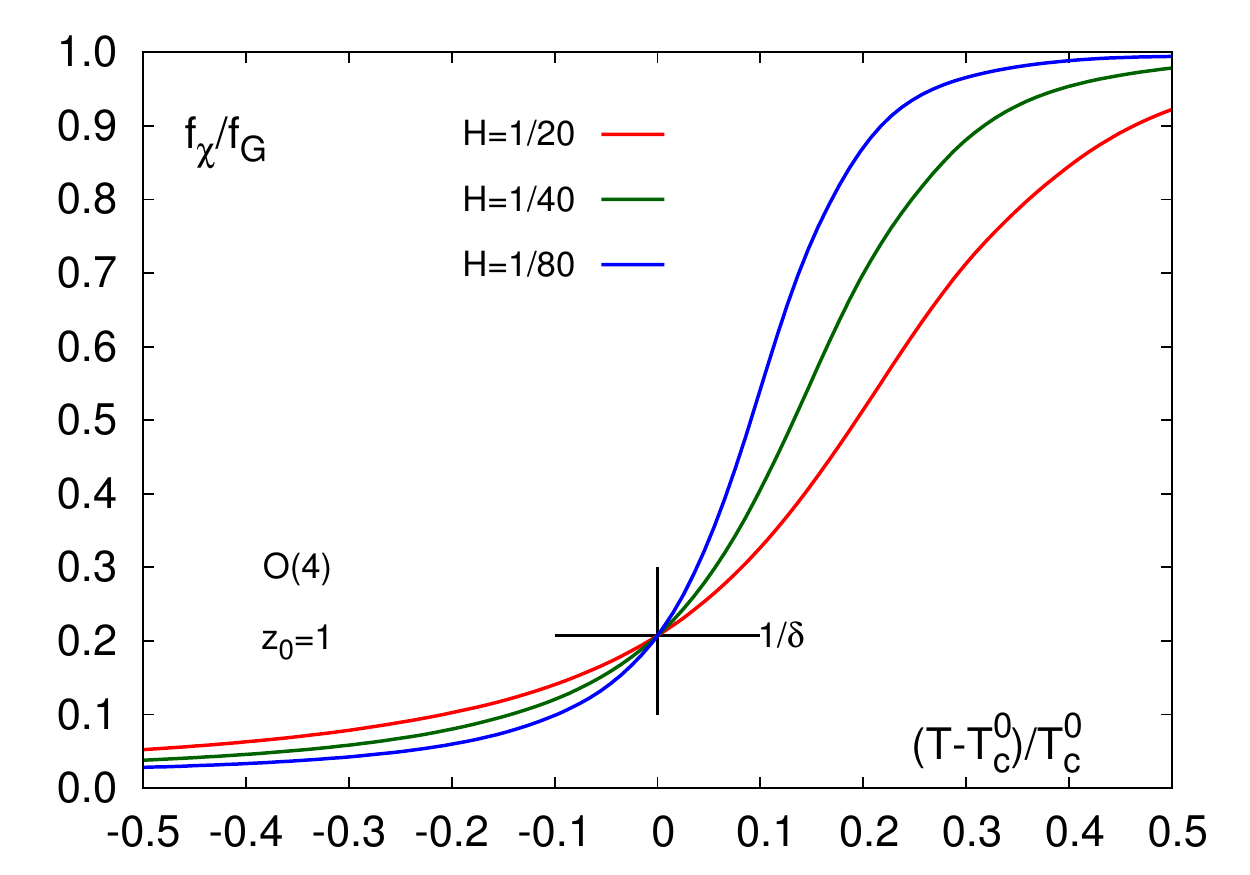}
\end{center}
\vskip -0.2 in
\caption{Left panel : Scaling function $f_{\chi}(z)$ for $O(4)$, $O(2)$ and $Z(2)$ universality class.
$z_p$ is the peak position which is universal. $z^-_{60\%}\approx 0$ for these universality classes.
Right panel : Ratio of scaling functions using $O(4)$ exponents for three different values of $H$.}
\label{fg.scaling}
\end{figure}

The scaling variable $z$ may be solved for $T$, which then gives the 
dependence of $T$ on $H$ when approaching the chiral limit at fixed $z$, 
\begin{equation}
T(z,H)=T_c^0\left( 1+\frac{z}{z_0}H^{1/\beta\delta} \right) .
\label{eq.Tc0fromfixedz}
\end{equation}
This is commonly used when analyzing the approach of the pseudo-critical temperature $T_{pc}$,
defined through the peak of $\chi_m$ located at $z=z_p$, to the chiral limit.
Here we consider a different choice of $z$. We propose to choose $z=z^-_{60\%}$;
the value of $z^-_{60\%}$ or equivalently $T^-_{60\%}$ is defined as
$\chi_{M}\left(T^-_{60\%}\right)=0.6\chi^{\rm max}_{M}$.
From the left panel of Fig.\ref{fg.scaling}, one can clearly see that for the scaling
functions, which are relevant for our discussion, $60\%$ of the peak corresponds to
$z \simeq 0$, {\it i.e.}\ to a temperature close to the critical point. The influence of
$H$-dependent corrections thus will be suppressed by at least an order of magnitude compared to $z_p$.
This in turn says that the estimator of $T_c^0$ using $T^-_{60\%}$ will be far more stable than that using $T_{pc}$.
\vskip -0.1 in
\vskip -0.1 in
Another estimate of the critical temperature can be obtained from the ratio $H\chi_{M}/{M}$
which behaves like a Binder cumulant at the critical point \cite{KL94} :
\begin{equation}
\lim_{H\to 0}\lim_{V\to\infty}\frac{H\chi_{M}\left(T_\delta,V,H\right)}{M\left(T_\delta,V,H\right)}=\frac{1}{\delta}
\Rightarrow T_c^0=\lim_{H\to 0}\lim_{V\to\infty}T_{\delta}(V,H) .
\label{eq.Tdelta}
\end{equation}
In the right panel of Fig.\ref{fg.scaling} we have shown this ratio using $O(4)$ scaling functions for three
different values of symmetry breaking parameter $H$. For simplicity we have set the scale $z_0=1$. From the
figure one can clearly see that in absence of corrections-to-scaling and 
regular terms the crossing point is unique for
different curves corresponding to different $H$ and has the value $1/\delta$.

To get some idea about the nature of the chiral transition we looked at the following ratio
\begin{equation}
\frac{M}{\chi_{M}}=\left(H-H_c\right)\frac{f_G(z)}{f_{\chi}(z)},
\label{eq.MbychiM}
\end{equation}
where $H_c$ is zero for $O(4)$ or $O(2)$ transitions and non-zero for $Z(2)$ transition.
It is clear from Eq.\ref{eq.MbychiM} that near the critical point the ratio will be linear in
$H$ and the slope is uniquely defined through the universal scaling functions. This means that one
can directly determine the slope and hence the universality class with precise enough
data. If the chiral transition is of $1^{\rm st}$ order then this ratio will have a sudden drop
at some non-zero value of $H_c$. One can also estimate this critical value $H_c$ from the ratio
using informations from $Z(2)$ universality class.

\vspace*{-0.5 cm}
\section{Results}
\label{sc.results}
\vspace*{-0.3 cm}

We have used the HISQ/tree action for numerical simulations of (2+1)-flavor QCD. To approach the chiral
limit we have decreased $m_l/m_s$ (keeping $m_s$ fixed at its physical value) corresponding to
55 MeV $\leq m_{\pi} \leq$ 160 MeV. We have used lattices with temporal extent $N_{\tau}=$6, 8, 12
and the spatial volumes used are in the range $4 \leq N_{\sigma}/N_{\tau} \leq 8$.
\vskip -0.1 in
\begin{figure}[H]
\begin{center}
\includegraphics[scale=0.50]{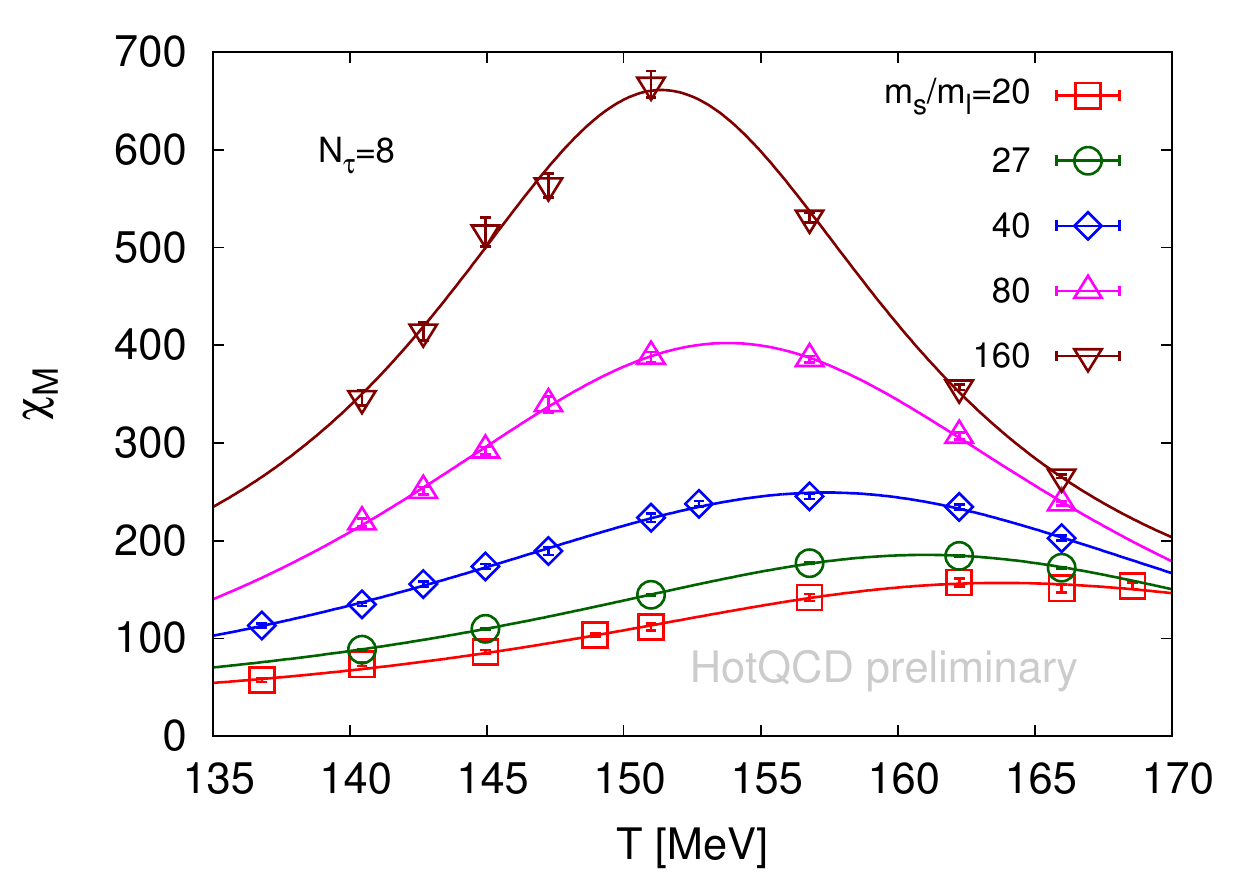} ~~~~~~~~
\includegraphics[scale=0.50]{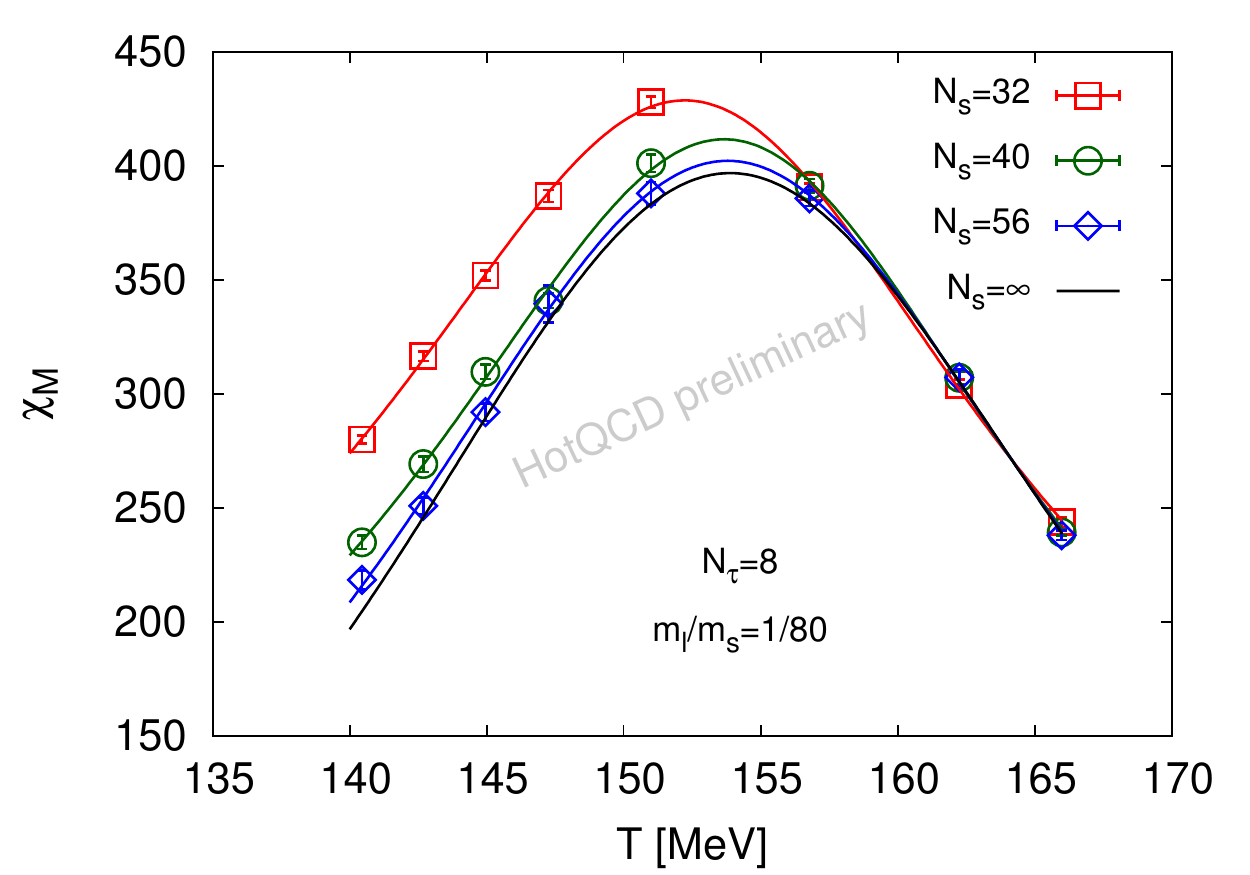}
\end{center}
\vskip -0.2 in
\caption{Left panel : $\chi_{M}$ for five different quark masses are plotted as a function
of $T$ for $N_{\tau}=8$. Right panel : $\chi_{M}$ vs.\ $T$ for three different spatial
volumes for $N_{\tau}=8$ with $m_s/m_l=80$.}
\label{fg.suscMapprox}
\end{figure}
\vskip -0.1 in
In the left panel of Fig.\ref{fg.suscMapprox} we have plotted $\chi_{M}$ for $N_{\tau}=8$ lattices
with five different quark masses. The increase of $\chi_{M}$ with decreasing quark mass is evident
and roughly consistent with the scaling expectations,
$\chi^{\rm max}_{M}\sim H^{1/\delta-1}$. In the right panel of Fig.\ref{fg.suscMapprox} we
have shown the volume dependence of $\chi_{M}$ for $m_s/m_l=80$ which is the next
to lowest mass for $N_{\tau}=8$ lattices. It is evident from the figure that
$\chi^{\rm max}_{M}$ decreases slightly as the volume increases, which is opposite
to what is expected for a $1^{\rm st}$ or $2^{\rm nd}$ order phase transition.
So we can eventually rule out the possibility of a $1^{\rm st}$ order phase transition
for $m_{\pi} \geq 80$ MeV. The black line in the plot comes from a linear extrapolation
in $1/V$ using the two largest volumes.
\vskip -0.1 in
\begin{figure}[H]
\begin{center}
\includegraphics[scale=0.50]{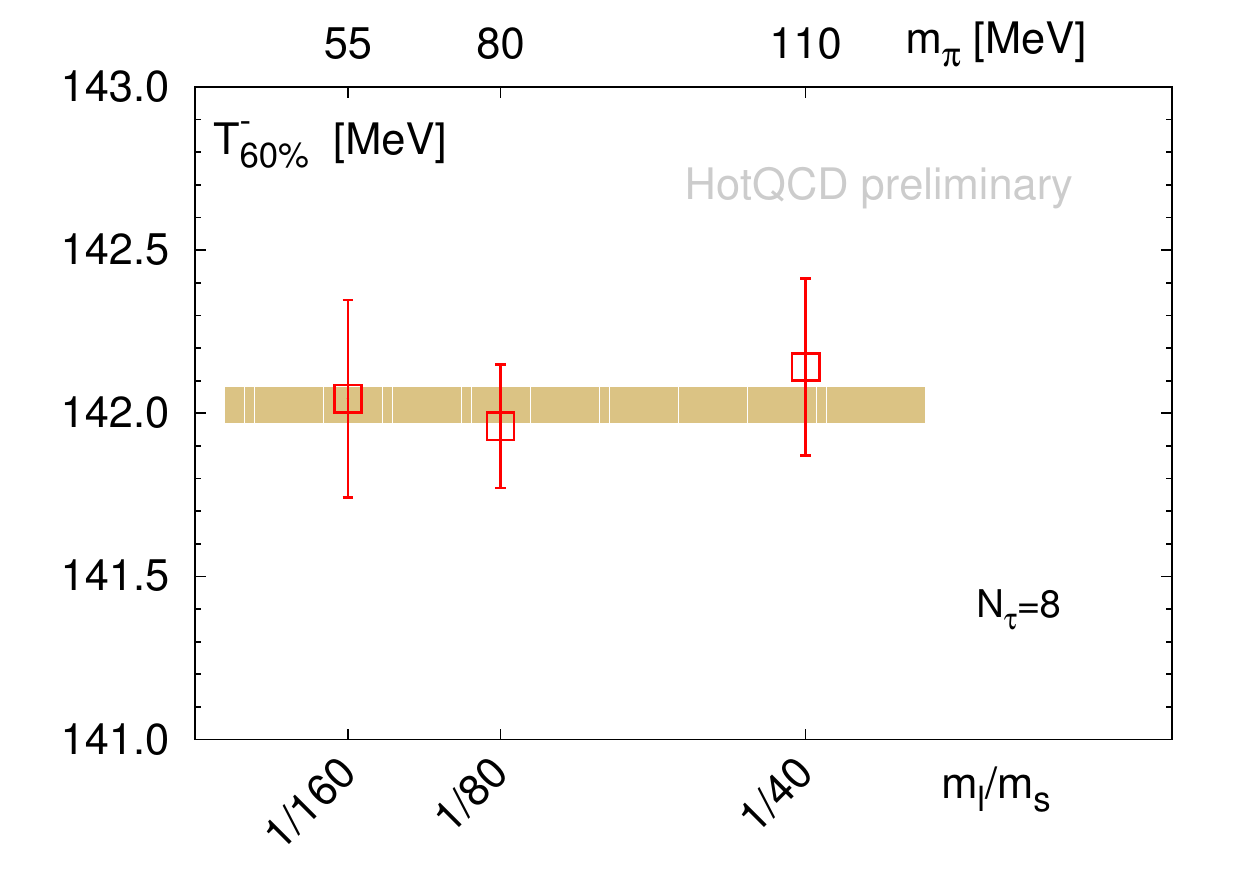} ~~~~~~~~
\includegraphics[scale=0.50]{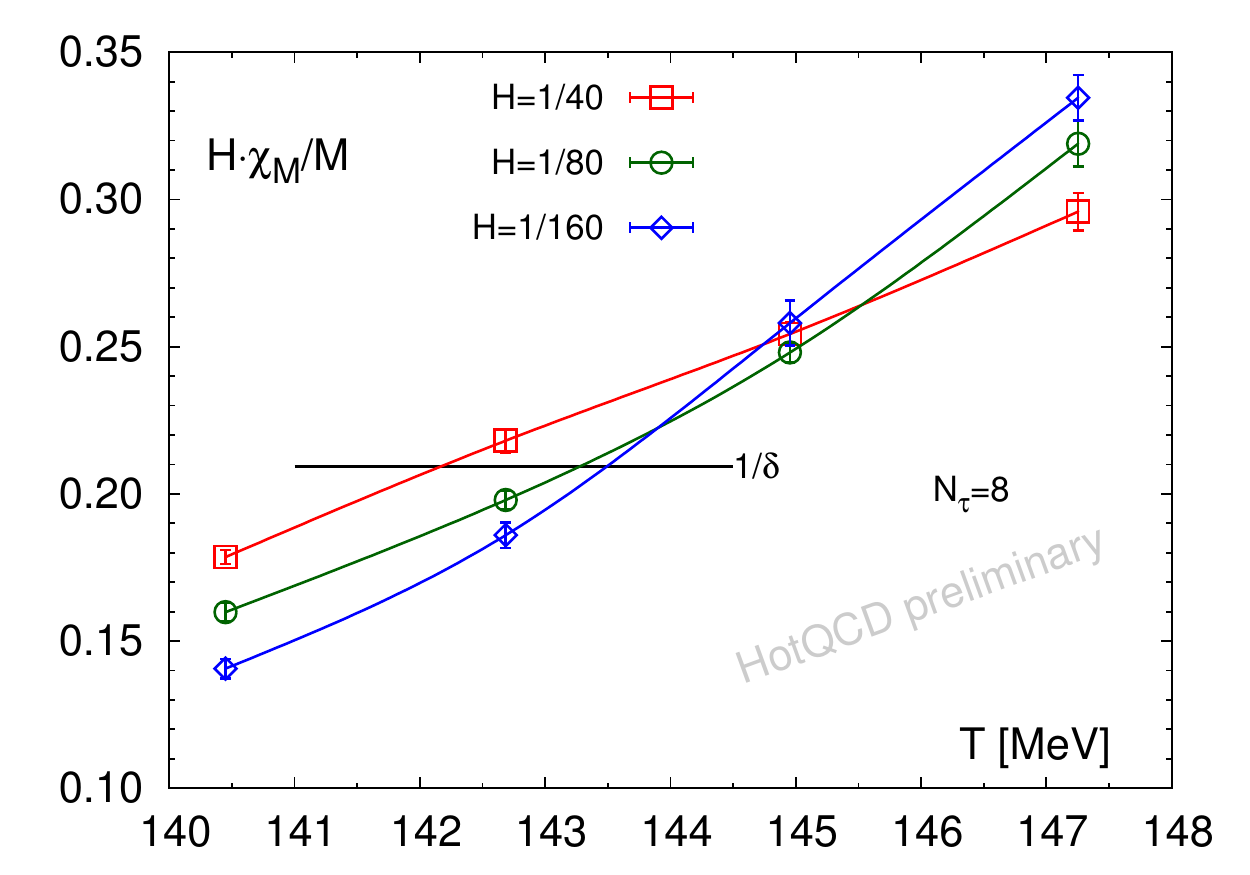}
\end{center}
\vskip -0.2 in
\caption{Left panel : $T^-_{60\%}$ vs.\ $H$ for $N_{\tau}=8$ lattices.
Right panel : $H\chi_{M}/{M}$ vs.\ $T$ for three lowest $H$ for $N_{\tau}=8$.}
\label{fg.T60Tdelta}
\end{figure}
\vskip -0.1 in
In the left panel of Fig.\ref{fg.T60Tdelta} we have shown $T^-_{60\%}$ calculated for $N_{\tau}=8$
lattices corresponding to the three lowest pion masses. As can be seen from the figure, $T^-_{60\%}$
is almost constant for low enough pion masses which is expected from Eq.\ref{eq.Tc0fromfixedz},
since $z^-_{60\%}/z_0\sim\mathcal{O}(10^{-2})$. This implies eventually fitting a constant to
$T^-_{60\%}$ in this regime can already give a reliable estimate of $T_c^0$. This constant is
around 142 MeV for $N_{\tau}=8$ which is shown along with its uncertainty by the band in the
figure. In the right panel of Fig.\ref{fg.T60Tdelta} we have shown the ratio $H\chi_{M}/{M}$
for three lowest pion masses for $N_{\tau}=8$ and the solid lines are splines to guide the eye.
The uniqueness of the crossing points, what we have discussed in Sec.\ref{sc.obsanddef} is
absent in the data. Our preliminary analysis suggests that deviations from this unique
crossing are mainly due to finite volume effects rather than contributions from regular
terms. A conservative estimate that takes these systematic effects into account, yields
$T_c^0\sim 144$ MeV for $N_\tau =8$. Performing a joint fit to $M$ and $\chi_{M}$ with
magnetic equation of state \cite{Ejirietal09}, a comparable value $T_c^0\sim$ 145 MeV
is obtained. Putting together all the above-mentioned estimates we arrive
at $T_c^0=$ 144(2) MeV which is the current estimate for $N_{\tau}=8$. Similar analyses
have also been carried out for $N_{\tau}=6$ and $N_{\tau}=12$ which gives the estimates
$T_c^0=$ 147(2) MeV and 139(3) MeV, respectively. A continuum extrapolation 
linear in $N_\tau^{-2}$, using results for different $N_{\tau}$, yields 
$$T_c^0=138(5) ~ {\rm MeV} ~ ({\rm continuum ~ HotQCD ~ preliminary}).$$
\vskip -0.2 in
\begin{figure}[H]
\begin{center}
\includegraphics[scale=0.50]{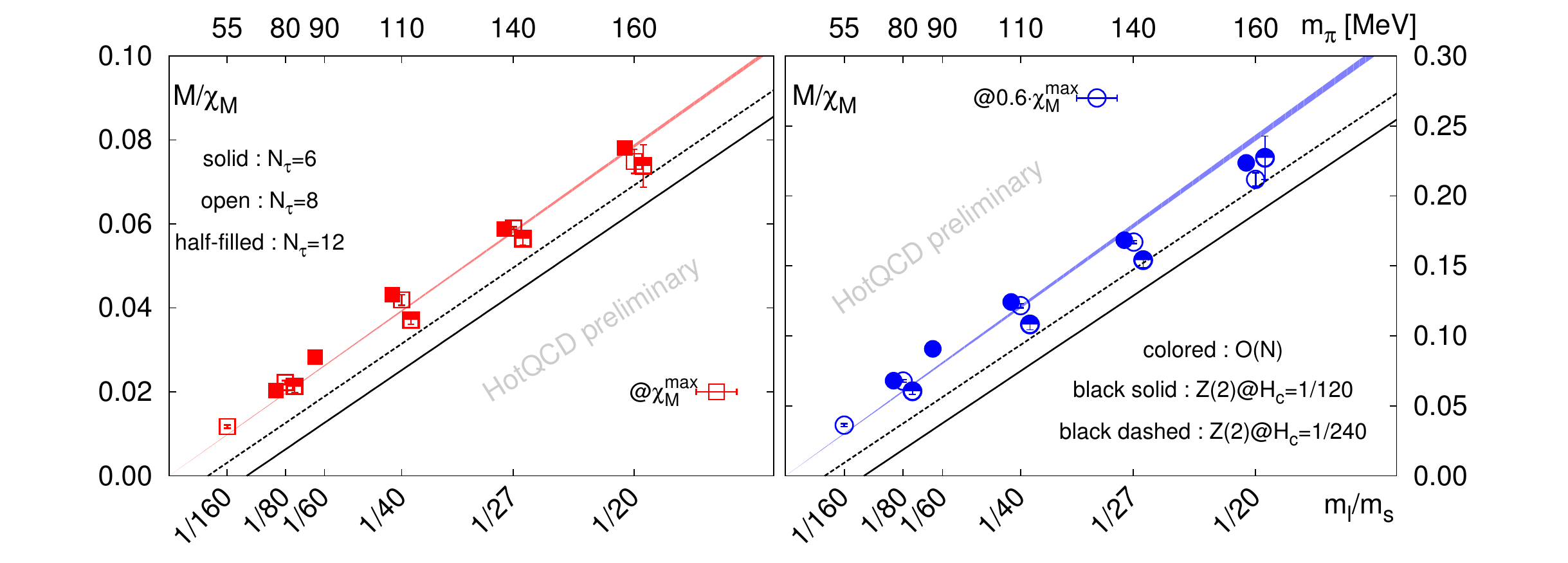}
\end{center}
\vskip -0.2 in
\caption{$M/\chi_{M}$ is plotted for different $N_{\tau}$ along with the scaling
expectations from different universality classes.}
\label{fg.MtochiMratio}
\end{figure}
Before concluding we would like to say a few (preliminary) words on the
order of the chiral phase transition and the corresponding universality class.
In Fig.\ref{fg.MtochiMratio} we have plotted $M/\chi_{M}$ as a function of $H$ for different
$N_{\tau}$ at two different positions. In the left panel, we have plotted the ratio at the peak
position of $\chi_{M}$ and in the right panel we have shown the same at the point where
$\chi_{M}$ attains 60\% of its maximum. For both the plots one can see for low enough masses
the ratio seems to behave linearly. The colored bands in the plots are not fits rather
expectations coming from $O(N)$ universality classes. The width in the band comes from
the small difference between $O(4)$ and $O(2)$. For a crude comparison we have also
plotted expectations from $Z(2)$ universality class by black lines for two different
values of $H_c$ : solid line is for $H_c=1/120$ and dashed line is for $H_c=1/240$.
The data seems to favor $O(N)$ expectations over $Z(2)$. Although one has to
keep in mind that for a non-vanishing $H_c$, $M$ is not any more an exact order parameter
and the $Z(2)$ lines in Fig.\ref{fg.MtochiMratio} will not be actually straight lines.

\vspace*{-0.5 cm}
\section{Conclusions}
\label{sc.conclusions}
\vspace*{-0.3 cm}

We have estimated the chiral critical temperature $T_c^0$ using (2+1)-flavor HISQ/tree action.
For $N_{\tau}=$ 6, 8 and 12 the preliminary estimates of $T_c^0$ are 147(2) MeV, 144(2) MeV
and 139(3) MeV, respectively. A continuum extrapolation using these numbers gives the
preliminary estimate that in continuum $T_c^0=$ 138(5) MeV. Our preliminary analyses seem 
to favor a $2^{\rm nd}$ order chiral phase transition over a $1^{\rm st}$ order transition.

\vspace*{-0.5 cm}
\section{Acknowledgments}
\label{sc.acknowledgments}
\vspace*{-0.3 cm}

This work was supported in part through contract No.\ DE-SC0012704 with the U.S.\ Department
of Energy, Scientific Discovery through Advance Computing (SciDAC) award \textquotedblleft Computing
the Properties of Matter with Leadership Computing Resources\textquotedblright, the Deutsche
Forschungsgemeinschaft (DFG) through the grant CRC-TR 211 \textquotedblleft Strong-interaction matter
under extreme conditions\textquotedblright, the grant 05P15PBCAA of the German
Bundesministerium f\"ur Bildung und Forschung, Early Career Research Award of the
Science and Engineering Research Board of the Government of India and the National Natural Science
Foundation of China under grant numbers 11535012 and 11775096.






\bibliographystyle{elsarticle-num}
\bibliography{<your-bib-database>}



\end{document}